\documentclass[twocolumn]{aastex631}
\usepackage{tablefootnote}
\usepackage[utf8]{inputenc}

\begin{document}
\maxdeadcycles=200

\def\H{$^1$H }
\def\geqsim{\lower.73ex\hbox{$\sim$}\llap{\raise.4ex\hbox{$>$}}$\,\,$}
\def\leqsim{\lower.73ex\hbox{$\sim$}\llap{\raise.4ex\hbox{$<$}}$\,\,$}

\newcommand{\LyA}{Lyman-$\alpha$ }
\newcommand{\Mgii}{\ion{Mg}{2} }
\newcommand{\CIV}{\ion{C}{4} }
\newcommand{\SiIV}{\ion{Si}{4} }
\newcommand{\Feii}{\ion{Fe}{2} }

\newcommand{\Mgiinospace}{\ion{Mg}{2}}
\newcommand{\CIVnospace}{\ion{C}{4}}
\newcommand{\SiIVnospace}{\ion{Si}{4}}
\newcommand{\Feiinospace}{\ion{Fe}{2}}

\newcommand{\ParentQSO}{83{,}207 }
\newcommand{\PISample}{374 }
\newcommand{\PrePICut}{29{,}797 }
\newcommand{\PostPICut}{29{,}423 }
\newcommand{\FinalSample}{23{,}921 }

\newcommand{\ADM}[1]{\textcolor{red}{\bf #1} \\} 
\newcommand{\LGN}[1]{\textcolor{blue}{\bf #1}} 

\title{Detecting and Characterizing \Mgii absorption in DESI Survey Validation Quasar Spectra}
\shorttitle{\Mgii Absorbers in DESI Quasar Spectra}

\author[0000-0002-5166-8671]{Lucas Napolitano}
\author{Agnesh Pandey}
\author{Adam D.\ Myers}
\affiliation{Department of Physics \& Astronomy, University of Wyoming, 1000 E. University, Dept. 3905, Laramie, WY 82071, USA}
\author[0000-0001-8857-7020]{Ting-Wen Lan}
\affiliation{Graduate Institute of Astrophysics and Department of Physics, National Taiwan University, No. 1, Sec. 4, Roosevelt Road, Taipei 10617, Taiwan}
\author{Abhijeet Anand}
\affiliation{Lawrence Berkeley National Laboratory, 1 Cyclotron Road, Berkeley, CA 94720, USA}
\author{Jessica Aguilar}
\affiliation{Lawrence Berkeley National Laboratory, 1 Cyclotron Road, Berkeley, CA 94720, USA}
\author[0000-0001-6098-7247]{Steven Ahlen}
\affiliation{Physics Dept., Boston University, 590 Commonwealth Avenue, Boston, MA 02215, USA}
\author[0000-0002-5896-6313]{David M. Alexander}
\affiliation{Centre for Extragalactic Astronomy, Department of Physics, Durham University, South Road, Durham, DH1 3LE, UK}
\author{David Brooks}
\affiliation{Department of Physics \& Astronomy, University College London, Gower Street, London, WC1E 6BT, UK}
\author{Rebecca Canning}
\affiliation{Institute of Cosmology \& Gravitation, University of Portsmouth, Dennis Sciama Building, Portsmouth, PO1 3FX, UK}
\author{Chiara Circosta}
\affiliation{Department of Physics \& Astronomy, University College London, Gower Street, London, WC1E 6BT, UK}
\author{Axel De La Macorra}
\affiliation{Instituto de Física, Universidad Nacional Autónoma de México, Cd. de México C.P. 04510, México}
\author{Peter Doel}
\affiliation{Department of Physics \& Astronomy, University College London, Gower Street, London, WC1E 6BT, UK}
\author{Sarah Eftekharzadeh}
\affiliation{Universities Space Research Association, NASA Ames Research Centre, USA}
\author[0000-0003-1251-532X]{Victoria A. Fawcett}
\affiliation{School of Mathematics, Statistics and Physics, Newcastle University, NE1 7RU, UK}
\affiliation{Centre for Extragalactic Astronomy, Department of Physics, Durham University, South Road, Durham, DH1 3LE, UK}
\author[0000-0002-3033-7312]{Andreu Font-Ribera}
\affiliation{Institut de Física d’Altes Energies (IFAE), The Barcelona Institute of Science and Technology, Campus UAB, E-08193 Bellaterra Barcelona, Spain}
\author{Juan Garcia-Bellido}
\affiliation{Instituto de Física Teórica (IFT) UAM/CSIC, Universidad Autónoma de Madrid, Cantoblanco, E-28049, Madrid, Spain }
\author[0000-0003-3142-233X]{Satya  Gontcho A Gontcho}
\affiliation{Lawrence Berkeley National Laboratory, 1 Cyclotron Road, Berkeley, CA 94720, USA}
\affiliation{Department of Physics and Astronomy, University of Rochester, 500 Joseph C. Wilson Boulevard, Rochester, NY 14627, USA }
\author[0000-0001-7178-8868]{L. Le Guillou}
\affiliation{Sorbonne Université, CNRS/IN2P3, Laboratoire de Physique Nucléaire et de Hautes Energies (LPNHE), F-75005 Paris, France}
\author{Julien Guy}
\affiliation{Lawrence Berkeley National Laboratory, 1 Cyclotron Road, Berkeley, CA 94720, USA}
\author{Klaus Honscheid}
\affiliation{Center for Cosmology and AstroParticle Physics, The Ohio State University, 191 West Woodruff Avenue, Columbus, OH 43210, USA}
\affiliation{Department of Physics, The Ohio State University, 191 West Woodruff Avenue, Columbus, OH 43210, USA}
\author{Stephanie Juneau}
\affiliation{NSF’s National Optical-Infrared Astronomy Research Laboratory, 950 N. Cherry Avenue, Tucson, AZ 85719, USA}
\author[0000-0003-3510-7134]{T. Kisner}
\affiliation{Lawrence Berkeley National Laboratory, 1 Cyclotron Road, Berkeley, CA 94720, USA}
\author[0000-0003-1838-8528]{Martin Landriau}
\affiliation{Lawrence Berkeley National Laboratory, 1 Cyclotron Road, Berkeley, CA 94720, USA}
\author[0000-0002-1125-7384]{Aaron M. Meisner}
\affiliation{NSF’s NOIRLab, 950 N. Cherry Ave., Tucson, AZ 85719, USA}
\author{Ramon Miquel}
\affiliation{Instituci\'o Catalana de Recerca i Estudis Avan\c{c}ats (ICREA), Pg.~de Llu\'{\i}s Companys 23, 08010 Barcelona, Spain}
\affiliation{ Institut de F\'{\i}sica d'Altes Energies (IFAE), The Barcelona Institute of Science and Technology, Campus UAB, 08193 Bellaterra Barcelona, Spain}
\author[0000-0002-2733-4559]{J. Moustakas}
\affiliation{Department of Physics and Astronomy, Siena College, 515 Loudon Road, Loudonville, NY 12211, USA }
\author[0000-0002-0644-5727]{Will J. Percival}
\affiliation{Department of Physics and Astronomy, University of Waterloo, 200 University Ave W, Waterloo, ON N2L 3G1, Canada}
\affiliation{Perimeter Institute for Theoretical Physics, 31 Caroline St. North, Waterloo, ON N2L 2Y5, Canada}
\affiliation{Waterloo Centre for Astrophysics, University of Waterloo, 200 University Ave W, Waterloo, ON N2L 3G1, Canada}
\author{J. Xavier Prochaska}
\affiliation{University of California, Santa Cruz, 1156 High Street, Santa Cruz, CA 95064, USA}
\affiliation{Kavli Institute for the Physics and Mathematics of the Universe, 5-1-5 Kashiwanoha, Kashiwa, 277-8583, Japan}
\author{Michael Schubnell}
\affiliation{Department of Physics, University of Michigan, 450 Church St., Ann Arbor, MI 48109, USA}
\author[0000-0003-1704-0781]{Gregory Tarlé}
\affiliation{Department of Physics, University of Michigan, 450 Church St., Ann Arbor, MI 48109, USA}
\author{B. A. Weaver}
\affiliation{NSF’s NOIRLab, 950 N. Cherry Ave., Tucson, AZ 85719, USA}
\author{Benjamin Weiner}
\affiliation{Steward Observatory, University of Arizona, 933 N, Cherry Avenue, Tucson, AZ 85721, USA}
\author[0000-0002-4135-0977]{Zhimin Zhou}
\affiliation{National Astronomical Observatories, Chinese Academy of Sciences, A20 Datun Rd., Chaoyang District, Beijing, 100012, P.R. China}
\author[0000-0002-6684-3997]{Hu Zou}
\affiliation{National Astronomical Observatories, Chinese Academy of Sciences, A20 Datun Rd., Chaoyang District, Beijing, 100012, P.R. China}
\author[0000-0002-3983-6484]{Siwei Zou}
\affiliation{Department of Astronomy, Tsinghua University, Beijing 100084, China}

\begin{abstract}
We present findings of the detection of Magnesium II (\Mgiinospace, $\lambda$ = 2796, 2803${\rm \AA}$) absorbers from the early data release of the Dark Energy Spectroscopic Instrument (DESI). DESI is projected to obtain spectroscopy of approximately 3 million quasars (QSOs), of which over 99\% are anticipated to be at redshifts greater than z $>$ 0.3, such that DESI would be able to observe an associated or intervening \Mgii absorber illuminated by the background QSO. We have developed an autonomous supplementary spectral pipeline that detects such systems through an initial line-fitting process and then confirms line properties using a Markov Chain Monte Carlo (MCMC) sampler. Based upon a visual inspection of resulting systems, we estimate that this sample has a purity greater than $99$\%. We have also investigated the completeness of our sample in regards to both the signal-to-noise properties of the input spectra and the rest-frame equivalent width (W$_0$) of the absorber systems. From a parent catalog containing \ParentQSO quasars, we detect a total of \FinalSample \Mgii absorption systems following a series of quality cuts. Extrapolating from this occurrence rate of 28.8\% implies a catalog at the completion of the five-year DESI survey that contains over eight hundred thousand \Mgii absorbers. The cataloging of these systems will enable significant further research as they carry information regarding circumgalactic medium environments, the distribution of intervening galaxies, and the growth of metallicity across the redshift range $0.3\leq z < 2.5$.

\end{abstract}


\section{Introduction}
In the years since the discovery of the first quasars \citep[e.g.][]{MS,Sch63}, these objects have become crucial cosmological tracers, helping to map the underlying mass distribution and history of structure formation across cosmic time \citep[e.g.][]{Cro05,SpringelQSO,She07,Ros09,She09,Whi12,Eft15,Zar18,Nev20}. Shortly after their discovery it was observed that the spectra of some quasars have absorption line systems at redshifts distinct from their emission. It was first proposed by \citet{Wagoner} and \citet{BahcallSpitzer} that these absorption lines may be caused by gas excitation in the extended halos, or circumgalactic medium (CGM), of intervening galaxies in the line of sight to the more distant quasar (see \citealt{CGM} for an overarching review of the study of the CGM). 

The host galaxies of these absorption systems are frequently too dim to be otherwise observed, particularly those at redshifts beyond $z=1$ \citep[e.g.][]{AbsHost1,AbsHost2,AbsHost3,AbsHost4}. However, the detectability of absorption systems associated with these hosts is not affected by the luminosity of the host system, nor by its redshift. As such, these absorption systems allow us to query the gas content of a diverse set of galactic environments that would otherwise be difficult or impossible to survey.

Although a range of absorption species are commonly found in quasar spectra, including Fe II, Al III, C IV, and Si IV, the analysis in this paper will focus on \Mgii as it produces a distinct doublet shape, and can be detected at rest-frame wavelengths between \Mgii emission near $2800\,{\rm \AA}$ and Ly$\alpha$ emission near $1215\,{\rm \AA}$. It becomes significantly more challenging to reliably detect absorption systems blueward of Ly$\alpha$ emission as they will often blend with lines in the Ly$\alpha$ Forest \citep[e.g.][]{MetalLya}.

The detection and characterization of \Mgii absorbers allows for the study of their distribution across redshift space. Since these absorbers are inherently associated with galaxies, they can be used as mass tracers and their density informs the formation of cosmic structure \citep[e.g.][]{Bergeron1991, ChenIncidence}. The determination of rest-frame equivalent width values can also provide insight into the physical scale of circumgalactic media, and their variation as a function of redshift \citep[e.g.][]{Bordoloi2011,Rajeshwari2023}. Additionally, as galaxies evolve the composition and physical properties of their CGM environments change, causing certain absorption species to become more or less detectable \citep[e.g.][]{Daddi,CGM}. Therefore, the detection of \Mgii absorbers can inform our understanding of galaxy evolution because only certain CGM environments can result in \Mgii absorption.

Absorption systems also enable studies of the nature and kinematics of outflowing gas in their host galaxies \citep[e.g.][]{Prochaska2004,Nestor2011,Bordoloi2011,Bouche2012,Kacprzak2012,Lan2018}, as well as the covering fractions and relative metallicities of the absorbing material \citep[e.g.][]{Steidel1994,Aracil2004,Chen2010,Lan2020}. By first detecting \Mgiinospace, it is possible to detect other metal lines arising from the same CGM and do so with greater certainty, even at lower EW values.

Numerous catalogs of \Mgii absorbers have been constructed dating back over forty years. These include those based on early, purpose-built surveys with samples of dozens to hundreds of systems \citep[e.g.][]{Lanzetta1987, Tytler1987, Sargent1988, Caulet1989, SteidelSargent1992, Churchill1999}, to high-resolution surveys with large telescopes \citep[e.g.][]{Nielsen2013, Chen2017}, to catalogs based on the large number of quasars available from the Sloan Digital Sky Survey (SDSS) \citep[e.g.][]{Nestor2005, York2006, Prochter2006, Lundgren2009, Quider2011, Zhu2013, Seyffert2013, Raghunathan2016, Siwei, Anand2021}. SDSS is responsible for the bulk of quasar observations to date, having observed $\sim$750{,}000 \citep[e.g.][]{QSOSDSS7,QSOSDSS12,EB}.

The Dark Energy Spectroscopic Instrument (DESI) began its Main Survey of the sky in May 2021. The DESI survey represents a significant improvement over the SDSS in terms of both raw data observed as well as data quality \citep{DESII}. DESI is mounted on the 4-meter Mayall telescope, which constitutes a roughly 250\% increase in light collecting area compared to the 2.5-meter telescope utilized in the SDSS \citep{DESIII}. DESI is also $\sim$160\% more efficient at passing light from the telescope to its spectrographs \citep{DESIII}. 

Over its five-year mission, DESI will observe approximately three million quasars \citep{QSOTS}. These quasars, along with DESI's other target classes, will enable studies of baryon acoustic oscillations and redshift-space distortions with the goal of determining new constraints on dark energy density and other cosmological parameters \cite{DESII}. With these objectives in mind DESI quasars can be divided into two classes, those observed as direct tracers ($z < 2.15$) and those observed in order to detect the foreground \LyA forest ($z > 2.15$)  \citep{SVPaper}.

DESI uses a combination of three optical bands (\textit{g,r,z}) as well as \textit{WISE} \textit{W1} and \textit{W2} band infrared photometry to select quasars based upon their infrared excess. The main quasar selection, detailed in \citet{QSOTS}, results in a selection of more than 200 deg$^{-2}$ quasars in the magnitude range $16.5 < r < 23$. Of these, approximately 70\% are projected to be direct tracer quasars, with the remaining 30\% being \LyA forest quasars.

DESI will have the capability to produce catalogs of absorbers that are significantly larger than those from any prior campaign. This will deepen our understanding of the innate distributions of \Mgii absorbers and their rest-frame equivalent widths, as well as enabling more precise analyses of galactic and CGM evolution \citep{Siwei2023}, structure formation \citep{PerezRafols} and relative metallicity evolution \citep{LanFukugita2017} than has previously been possible.

This paper is organized as follows. In Section 2, we will describe our techniques for detecting absorption systems, as well as our methods for determining the purity and completeness of the absorber sample. In Section 3, we will present our results, including the redshift distribution of detected systems and their rest-frame equivalent widths. In Section 4, we will discuss some possible applications of our absorber catalog --- in particular as a novel check on DESI pipeline redshifts and as a marker that can be used to locate other species in absorption. We present our conclusions in \S5.

\section{Data and Methods}
In this section we will describe the nature of DESI data and the construction of our catalog. We will then discuss how we have estimated the purity and completeness of our sample through the combined use of visual inspection, reanalysis of individual observations and simulations of \Mgii absorbers.
\subsection{DESI Data Construction}
\label{sec:Data}
DESI spectra are observed using three spectrographs, commonly referred to as ``b'', ``r'', and ``z'' due to their wavelength coverage, that together span a wavelength region of $3600\,{\rm \AA}$ to $9824\,{\rm \AA}$. The three spectrographs have approximately constant wavelength resolutions of $\Delta \lambda \approx 1.7 \rm \AA$ \citep{DESIII,DESIInstrument}. The flux values associated with a DESI observation are extracted using a linear wavelength grid in $0.8\,{\rm \AA}$ wavelength steps \citep{PipelinePaper}.

In this paper, we will use data from DESI's Early Data Release (EDR). These data are separated into three stages of survey validation, which we refer to as ``sv1'', ``sv2'', and ``sv3''. These surveys are distinct from each other both in time frame and targeting implementation and are detailed in \citet{TargetingPaper} and \citet{SVPaper}. Each of these stages, as well as the Main Survey, are further divided into bright-time and dark-time programs.

The typical effective exposure time DESI aims to achieve during dark-time observations is 1000 seconds. This timing is set by the need to classify and obtain redshifts for the Luminous Red Galaxy sample that is observed concurrently with the quasar sample during dark-time \citep[see Section 4 of][]{DESILRG}. The relationship between effective exposure time and signal-to-noise is detailed in Section 4.1.4 of \citet{PipelinePaper}. Note that quasars are only intended to be targeted during dark-time, see \citet{TargetingPaper} for further discussion of the distinction between DESI's bright-time and dark-time programs.

DESI will, over the course of its survey, re-observe targets to improve the quality of its data. During the Main Survey \LyA forest quasars are scheduled, at high priority, to be observed with four times the exposure time at which direct tracers quasars are typically observed, although all quasars can ultimately be re-observed at low priority \citep[see section 5.3.3 of][for details]{Schlafly}. During survey validation all quasars at redshifts $z > 1.6$ were observed for four times the typical exposure time \citep[e.g.][]{QSOTS}. This distinction carries implications for our completeness with respect to redshift, we comment on this further in \S\ref{sec:SampComplete}.

To perform our search for absorbers in the most robust fashion, we chose to use spectra that coadd all observations of a given target. We will refer to these spectra as being ``healpix coadded'' as this is how they are grouped within the DESI spectral reduction file structure \citep{Healpix}. We will also make use of the spectra that coadd a subset of observations, i.e.,\ all observations from a single night, in order to determine the completeness of our sample (see \S\ref{sec:SampComplete}).

Observations of the same target made during different survey validation stages are not coadded, and as such it is possible that a target could be present in multiple of these surveys. In such cases we will only consider the results of our absorber search for only the healpix coadded spectrum from the survey which has the highest squared template signal-to-noise (TSNR2; see section 4.14 of \citealt{PipelinePaper} for a full description of this statistic).

TSNR2 is calculated for different target classes (i.e. emission line galaxies, luminous red galaxies, quasars) according to their expected spectral properties and redshift distribution. This results in a more informed statistic that better weights relevant spectral features such as emission lines and the Lyman-$\alpha$ Forest. Notably, TSNR2 values for different target classes cannot be fairly compared, so when we refer to TSNR2 generically throughout this paper we will be referring to the TSNR2\_QSO statistic.

\subsection{Pipeline Construction}
\label{sec:pipeconstruct}
Our analysis relies on parent quasar catalogs generated via three tools: {\tt Redrock} (RR) \citep{RedRockPaper}\footnote{\url{https://github.com/desihub/redrock}}, a PCA-based template classifier that is part of the main DESI spectroscopic pipeline \citep{PipelinePaper}; {\tt QuasarNet} (QN) \citep{QN}\footnote{\url{https://github.com/ngbusca/QuasarNET}}, a neural network based quasar classifier \citep[see also][]{Farr2020}; and an \Mgiinospace-emission-based code designed to identify AGN-like spectra that show both strong galactic emission features and broad \Mgii emission. Results from the visual inspection of quasar spectra informed the need for, and use of, these tools \citep{QSOVI}.

Initial spectral types (QSO or non-QSO for our purposes) as well as initial redshifts are determined by RR. QN and the \Mgiinospace-emission code are then run as afterburners. The outputs of the two afterburners can result either in RR being re-run with adjusted redshift priors, or in the case of the \Mgiinospace-emission code the spectral type being changed to QSO when a broad \Mgiinospace-emission line is detected. Notably, redshifts are always ultimately determined by RR. A more complete overview of the application of these tools to construct quasar catalogs, as well as the verification of the completeness and purity of this approach can be found in \citet{QSOTS}. The resulting catalog has a purity greater than $> 99\%$ and a median redshift of $z=1.72$, with 68\% of quasars having redshifts between $1.07<z<2.46$ \citep{QSOTS}.

\begin{figure}[htb!]
    \epsscale{1.15}
    \plotone{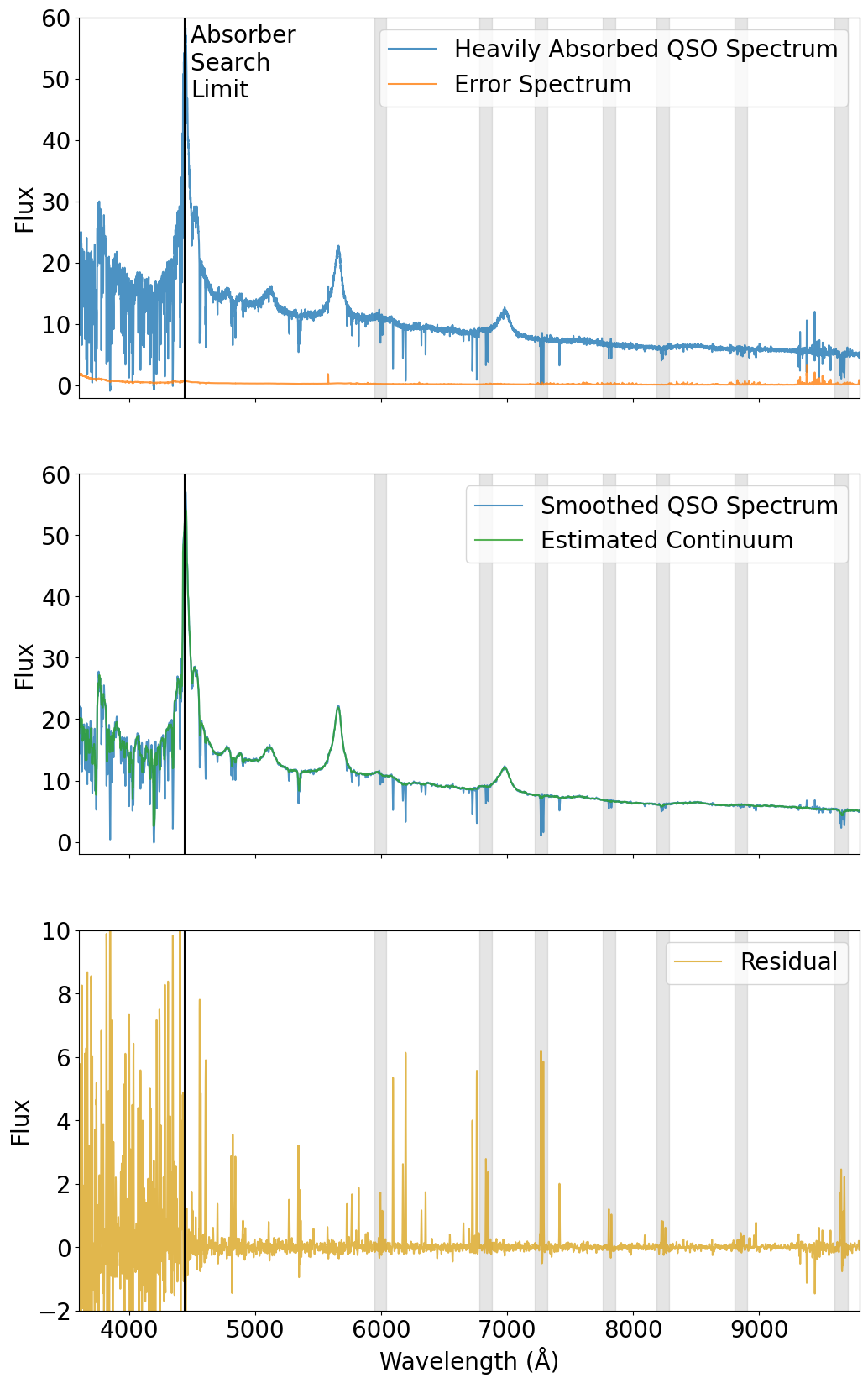}
    \caption{A visualization of the doublet-detection step of our pipeline. Detected \Mgii systems are shown in the gray outlined boxes and the search limit of our approach is shown by the vertical black line. \textit{Top}: A sample spectrum from DESI that features 7 separate \Mgii absorption systems. Shown are the flux and error spectrum coadded at the boundaries of the three DESI spectrographs. \textit{Middle}: The same spectrum now shown with an applied Gaussian smoothing kernel (blue) and estimated median-filter continuum (green). \textit{Bottom}: A residual taken from subtracting the median-filter-estimated continuum from the Gaussian-smoothed data. Note that the seven \Mgii absorption systems appear as positive lines.}
    \label{ExPipeline}
\end{figure}

We search these spectra for absorption doublets by first applying a Gaussian smoothing kernel with a standard deviation of two, as described in the {\tt Astropy} documentation \citep{Astropy}. This smoothing is performed in order to reduce the effect of noisy regions/pixels on the background quasar continuum we estimate in the following step. We estimate this continuum from the smoothed flux values of the spectra using a combination of median filters. Specifically we choose to weight the combination of a nineteen and thirty-nine pixel filter such that the contribution of the narrower, nineteen pixel filter, is strongest at low wavelengths, and decreases across the wavelength space, whereas the opposite is true for the thirty-nine pixel filter which contributes to the estimated continuum value most strongly at high wavelengths. 

These pixel values were informed by a preliminary set of \Mgii absorption detected using more rudimentary methods. The precise values have been chosen to ensure that the two absorption lines of the \Mgii doublet are cleanly separated into individual absorption lines by the estimated continuum. We find that this combination reliably models the broad emission features observed in DESI quasar spectra, but not any narrow absorption features that may be present. The choice to effectively broaden the filter at high wavelengths accounts for the broadening of \Mgii systems due to redshift.

An example of both the smoothing process and the estimated continuum can be seen in the central panel of Figure \ref{ExPipeline}. The bottom panel of Figure \ref{ExPipeline} demonstrates that emission features are not retained in a residual obtained by subtracting the median-filter-estimated continuum from the Gaussian-smoothed data, but any absorption features remain as positive features in the residual. Also evident is the increased residual noise beyond our search limit in the \LyA forest, making evident why it is difficult to resolve metal lines in this region.

To detect doublets we find every group of consecutive positive residuals and calculate a signal-to-noise ratio as:
\begin{equation}
    S/N = \frac {\sum_{p1}^{p2} C-F} {(\sum_{p1}^{p2} \sigma^2)^{1/2}}
\end{equation}

\noindent where $C$ and $F$ are the continuum and flux values respectively, $\sigma^2$ is the variance of the spectrum, and $p1$ and $p2$ are the first and last indices of a particular group of consecutive positive residuals. We additionally fit a preliminary Gaussian model to each set of residuals, this allows us to estimate the line center, and the determined values for line amplitude and width are later used to inform the initial state of the MCMC sampler. If two absorption lines are found that have SNR values greater than 2.5 and 1.5 respectively and a rest-frame wavelength separation of $7.1772\,\pm\,1.5$\,\AA, where 7.1772\,\AA\ is the laboratory separation of \Mgii \citep[e.g.][]{MgIIWave}, this doublet is regarded as being a likely candidate. 

A similar detection method was used in \citet{Raghunathan2016}, however we find that our approach improves the detection of relatively low signal absorption systems in high-signal QSO spectra. Note that our rest-frame line separation uncertainty value, $1.5$\,\AA, has been chosen in order to consider as many candidate absorbers as possible without encompassing the rest-frame separation of \SiIVnospace, another absorber doublet that is commonly strong in QSO spectra.  

To further verify these systems, we next perform an MCMC analysis using the {\tt emcee} software \citep{Emcee}. The decision to use MCMC to fit the relatively simple model of a doublet absorption line was made in the interest of fully understanding the posterior distributions of our parameters, as well as increasing the likelihood of recovering low-signal absorbers. 

To ensure that the full signal of the absorber is recovered we use a different continuum in fitting the systems than the one previously described and used in the initial detection step. The continuum used in detection is designed to ensure that the individual lines of the \Mgii doublet are detected separately, in the case of particularly strong or broad absorbers this can result in some of the absorption signal being lost in the residual. However, when fitting the absorber we instead construct a continuum to ensure that the the full signal is retained.

We first attempt to calculate an appropriate QSO continuum using the {\tt NonnegMFPy} tool as implemented in \citet{Zhu2013} and \citet{Anand2022}. {\tt NonnegMFPy} utilizes nonnegative matrix factorization (NMF, see \citealt{NMF}) to determine a basis set of eigen-spectra and through their reconstruction estimate an observed quasar continuum.

In cases where the NMF tool is unable to estimate a continuum due to an inability to converge, or the chi-squared value of the estimated continuum is greater than 4 (approximately thirty-two percent of DESI EDR QSOs) we estimate a secondary continuum using a wide, 85 pixel median-filter. In cases where this median-filter continuum provides a spectra fit with a lower chi-squared value we instead use this continuum. The width of this median filter is informed by the previously referenced preliminary sample of detected \Mgii absorbers and ensures that no signal is lost in estimating the continuum, even for the broadest absorption systems.

For each absorber candidate we consider a region of 80 pixels, or 64\,\AA, around the detected doublet; this value allows the sampler to explore a region of redshift space which will ultimately be much larger than the redshift uncertainty for a high quality fit, while simultaneously allowing for the detection of multiple \Mgii systems in a single spectrum. We then fit a five-parameter model of the form:
\begin{equation}
    F = A_1\exp{\frac{-[\lambda-C_{1}]^2}{2\sigma_1^2}}+ A_2\exp{\frac{-[\lambda-C_{2}]^2}{2\sigma_2^2}}
\end{equation}

\noindent where the two Gaussian line profiles are defined by their center $C$, width $\sigma$ and amplitude $A$. Note that $C_1$ and $C_2$ are both set by the same underlying redshift parameter, i.e. $C_1=(z+1)\times2796.3543$\,\AA.

The only prior attached to this model is the redshift range implied by the 80-pixel region around the suspected doublet. Initial values for our parameters are informed by the results of the detection step, with the minimum value in the group of residuals informing the line amplitude and the number of consecutive negative pixels informing the standard deviation. 

We use 32 walkers and run the model for 15{,}000 steps. We then discard the first 1000 steps as a burn-in period and store the remaining 14{,}000 steps for each candidate MCMC feature. Finally, we select only those models which have high mean acceptance fractions ($>0.45$) and estimated integrated autocorrelation times\footnote{See \url{https://emcee.readthedocs.io/en/stable/tutorials/autocorr/}} that are less than 1~per~cent of the chain length, indicating that the majority of proposed steps were accepted and that the model was well fit by the MCMC process. 

%These figures removed at this time
%A comparison between a model with a suitable acceptance fraction and autocorrelation estimates and a model with an unsuitable acceptance fraction can be seen in Figure \ref{ChainCompare}. 

%A further visualization of the outcome of a successful \Mgii fit is shown in Figure \ref{CornerPlot} in the Appendix, depicting the corner plot and generated model parameters for a detection.

After running this pipeline on our parent sample of \ParentQSO DESI QSOs we find a total of \PrePICut systems in 18{,}219 individual healpix coadded spectra that meet our criteria. 

In this sample there are a small number of entries with \Mgii absorption redshifts greater than the background quasar redshift, which may initially seem to suggest that the absorber is more distant the quasar,\footnote{Note that quasar redshifts are commonly determined using broad emission features, which naturally have a higher uncertainty than the redshifts determined using narrow absorption features.} In analyzing the physical interpretation of this scenario, it is standard to work in velocity rather than redshift space, and to define a velocity offset as:
\begin{equation}
    v_{\rm off} = c \frac{z_{\rm MgII}-z_{\rm QSO}}{1+z_{\rm QSO}}
\end{equation}
Velocity offset values within approximately $\pm6000$ ${\rm km\,s}^{-1}$ are indicative of an associated absorption system, wherein the QSO emission and metal line absorption arise from the same galaxy, or galaxy cluster \citep{ShenMenard2012}. However, in systems with larger velocity offset values it must be true that one of the redshifts is poorly determined --- as it is physically impossible for a system that is absorbing the light from a quasar to lie {\em behind} that quasar.

From a brief visual inspection of these systems we find a number of true \Mgii systems in quasar spectra with incorrect redshifts. Additionally, we find a number of false \Mgii systems; these are often detected in star or galaxy spectra that have been misidentified as quasars, or spectra with unusual error features. We therefore decide to group those systems with velocity offsets greater than 5000\,${\rm km\,s}^{-1}$ into a separate catalog of physically impossible absorbers for the purpose of diagnosing spectra that have been misclassified as quasars, or assigned an incorrect redshift. We will comment on this separate catalog further in \S \ref{sec:PIAs}. Removing the \PISample entries with $v_{\rm off} >$ 5000\,${\rm km\,s}^{-1}$ results in a preliminary sample of \PostPICut suspected \Mgii absorbers.

\subsection{Visual Inspection}
\label{sec:VI}
In the interest of both assessing and potentially improving our catalog purity we next conducted a visual inspection of 1000 randomly selected systems that pass the MCMC process, and do not have physically impossible redshifts, as described above. This process involves multiple steps: confirming that the background spectrum is indeed that of a quasar, verifying that two absorption lines have been well fit by the MCMC process, and determining if additional metal lines can be fit at the same redshift. Note that the presence of additional metal lines is considered only in confirming borderline cases where the fit absorption lines are weak.

Visually inspecting 1000 randomly selected \Mgii absorbers, following the steps outlined above, we find 808 which constitute true \Mgii absorption and 192 which do not. This suggests an initial purity of 80.8\%. Based upon the statistics of these systems we have developed a series of quality cuts to improve the purity of our detected sample with minimal effect on completeness.

The first cut we perform is to remove systems for which one or both of the fit Gaussians have a positive amplitude. This outcome is not disallowed by the MCMC priors to facilitate a full exploration of the parameter space, but is clearly not indicative of an absorption feature. Note that in such a case the initial line amplitude values were given as negative, however in the course of fitting the MCMC process has converged on a positive line solution. There are seventy-seven such systems in the visual inspection set, all of which were identified as false \Mgii systems and as such we impose a cut that all systems are required to have negative amplitudes for both fit line profiles.

\begin{figure}[htb!]
    \epsscale{1.15}
    \plotone{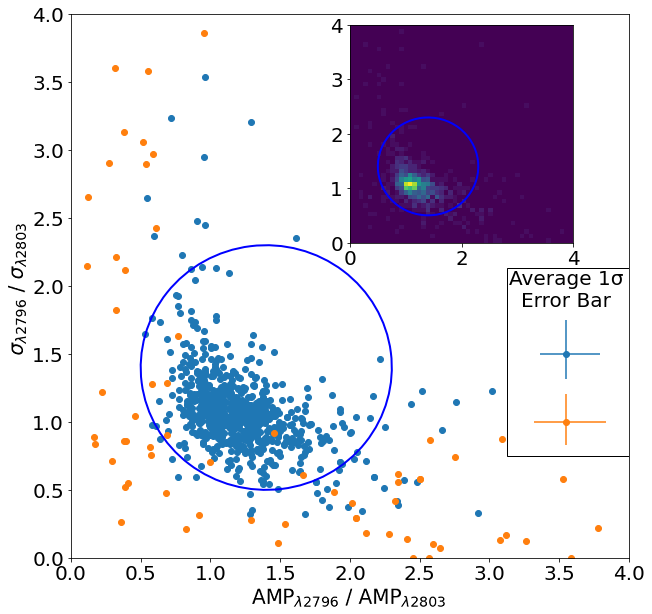}
    \caption{Visualization of Eqn.\,\ref{eqn:lineselect}. Plotted are the ratios between the widths and amplitudes of the two lines of the \Mgii doublet for all systems in the visually inspected set following the removal of any systems with positive amplitudes. True \Mgii systems are shown in blue and false systems are shown in orange. Note that not all points are shown. \textit{Top-Right Inset}: Density plot indicating that the distribution is highly concentrated around [1.1, 1.1].}
    \label{LineSimilarity}
\end{figure}

We can next consider another class of possible contaminants, systems in which two Gaussians with negative amplitude can be fit at the proper separation of \Mgiinospace, but whose line amplitudes and/or widths are not characteristic of \Mgiinospace. In order to determine an appropriate selection we must consider both the physical nature of \Mgii absorption, as well as the Gaussian line parameter posteriors of our visually inspected sample. To visualize these posteriors we have plotted the ratio of line amplitudes against the ratio of line widths in Figure \ref{LineSimilarity}. In both cases we consider the statistic of the leading 2796$\textrm{\AA}$ line divided by the statistic of the 2803$\textrm{\AA}$  line.

From the inset panel of this Figure we observe that the distribution of our visual inspection set in this space is tightly clustered around a value of roughly [1.1, 1.1], indicating (empirically) that the 2796$\textrm{\AA}$ line of the \Mgii doublet tends to have a slightly larger amplitude and be slightly wider than the second. 

The innate flux ratio of the 2796 to 2803 \AA\ lines is determined by the ratio of their collisional rate coefficients or equivalently by their quantum degeneracy factors. This ratio for \Mgii is F$_{2796}$/F$_{2803}$ = 2, and has been experimentally verified \citep[e.g.][]{MgII_I,MgII_II}. However, as the majority of systems observed here are saturated, the observed ratio of absorption line area approaches 1. 

In visually inspecting these systems we observe some true \Mgii absorbers where the amplitude and/or width of the 2803$\textrm{\AA}$ line is greater than that of the 2796$\textrm{\AA}$ line. This should not be possible theoretically, however the systematic uncertainties inherent to observation can produce this result. With this in mind we can draw a selection in this parameter space that includes the region of highest density / physical likelihood and allows for slight variation due to observational uncertainties, while still maximizing the purity of the post-cut selection. The selection takes the form of a circle and is described by:

%\begin{equation}
%     \label{eqn:lineselect}
%     (x-1.4)^2+(y-1.4)^2<0.81
%\end{equation}

\begin{equation}
     \label{eqn:lineselect}
     \left(\frac{{\rm AMP}_{\lambda2796}}{{\rm AMP}_{\lambda2803}}-1.4\right)^2+\left(\frac{\sigma_{\lambda2796}}{\sigma_{\lambda2803}}-1.4\right)^2<0.81
\end{equation}

%\noindent where $x$ and $y$ are the ratios of line amplitudes and widths respectively. Note that all points with amplitude and width ratios between 1.0 and 2.0 are included in this selection. After applying a cut to our sample according to the boundaries of this circle, we remove fifty true positives and one-hundred-eight false positives.

\noindent where ``AMP" is the amplitude of the fit line and $\sigma$ the width. Note that all points with amplitude and width ratios between 1.0 and 2.0 are included in this selection. After applying a cut to our sample according to the boundaries of this circle, we remove fifty true positives and one-hundred-eight false positives.

After imposing these cuts, the visual inspection sample contains 758 true positive \Mgii systems and 7 false positives for a nominal 99.1\% purity. Presuming Gaussian noise on this measurement we assign a $\pm 3.16 \%$ error on this purity. Applying these cuts to our pre-visual inspection sample of \PostPICut absorber candidates leaves a population of \FinalSample absorbers.

\subsection{Sample Completeness}
\label{sec:SampComplete}
In this section we will investigate the completeness of our sample. First we will determine its dependence on the quality of input spectra by using nightly data reductions. We then consider its relationship with the rest-frame equivalent width of absorption systems as determined by the creation of simulated absorbers.
\subsubsection{Reanalysis of Nightly Reduction}
\label{sec:Complete}

The healpix coadded spectra that we search for absorbers are created by combining multiple individual observations, as described in \S\ref{sec:pipeconstruct}. These individual observations have S/Ns, which makes recovering absorption features more challenging. This allows for a natural test of the completeness of our approach with respect to the TSNR2 of input spectra.

By searching the nightly coadded spectra, generally composed of a few individual observations, for a known \Mgii absorber, we can quantify the performance of our pipeline as a function of absorber redshift and spectral S/N. We choose to use the nightly coadded spectra rather than individual observations as this spans the region of relevant TSNR2 values more fully. Additionally, in rare cases where a target is observed on only a single night we do not consider the results of its reanalysis as the healpix coadded and nightly coadded spectra are the same. 

\begin{figure}[htb!]
    \epsscale{1.15}
    \plotone{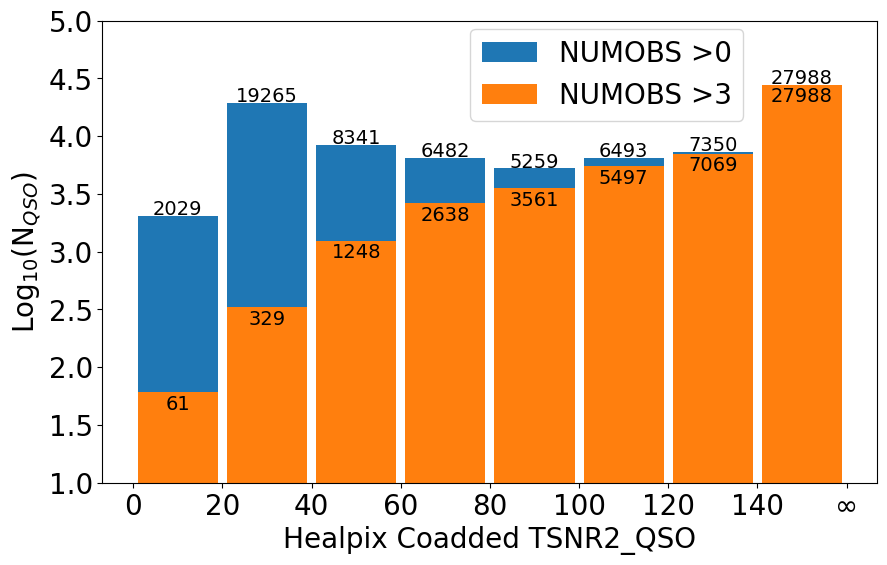}
    \caption{The population of healpix coadded quasar spectra TSNR2 values. The numbers above/below the blue/orange bars indicate population size. The right-most bin includes all spectra with TSNR2$>$140.}
    \label{QSO_pop}
\end{figure}

Figure \ref{QSO_pop} shows the TSNR2 distribution of healpix coadded quasar spectra. Results are shown for both targets with any number of observations and those targets with at least 4 observations. Note that we have grouped quasars with a TSNR2 value $>140$ as above this threshold we find that \Mgii detection is not sensitive to the TSNR2 of the background quasar. Additionally, we note that this final bin happens to contain only entries with at least four observations and accounts for approximately one third of the full healpix coadded sample.

\begin{figure}[htb!]
    \epsscale{1.15}
    \plotone{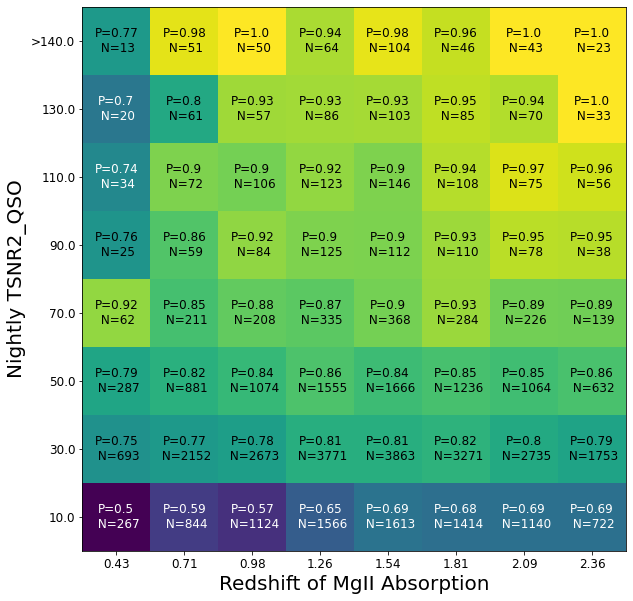}
    \caption{Nightly coadded QSO spectra grouped by \Mgii absorption redshift and TSNR2 value. Bins are colored according to the completeness, i.e., the number of \Mgii doublets recovered compared to the number of expected absorbers. Within each bin the completeness is quoted as P and the number of expected absorbers is quoted as N. Note that, as in Figure~\ref{QSO_pop}, the highest bin groups all spectra with a TSNR2 exceeding 140.}
    \label{CompleteGrid}
\end{figure}

Having determined the population of TSNR2 values we can now determine the performance of our pipeline in recovering known \Mgii absorbers as a function of TSNR2. In order to do so we consider all \FinalSample detected absorbers and recover the spectra of their nightly coadded observations. We then run the doublet-finder portion of our pipeline on these observations, recording whether the known \Mgii doublet can be recovered in these lower-TSNR2 spectra. The results of this search are displayed in Figure~\ref{CompleteGrid} --- as can be readily seen our percentage of recovered absorbers decreases with TSNR2, as anticipated. The percentage of recovered absorbers is also noticeably worse in the lowest redshift bin; this is because the DESI instrument has lower throughput at the blue end \citep{DESIInstrument}.

We can next consider the average completeness per TSNR2 bin, averaging across redshift-space. By multiplying the number of quasars in each bin of TSNR2 (i.e.\ Figure \ref{QSO_pop}), by this average completeness per bin we can determine the number of quasars in each bin for which we would expect to be able to recover \Mgii absorbers. Summing these results across TSNR2 bins and normalizing by the total number of quasars we determine the expected completeness. Doing so for the results shown in Figure~\ref{CompleteGrid} we recover an expected completeness of 88.0\% for QSOs with any number of observations and 92.8\% for QSOs with at least four observations.

\subsubsection{Injection of Synthetic Absorbers}
\label{sec:SynSim}

\begin{figure*}[htb!]
    \epsscale{1.0}
    \plotone{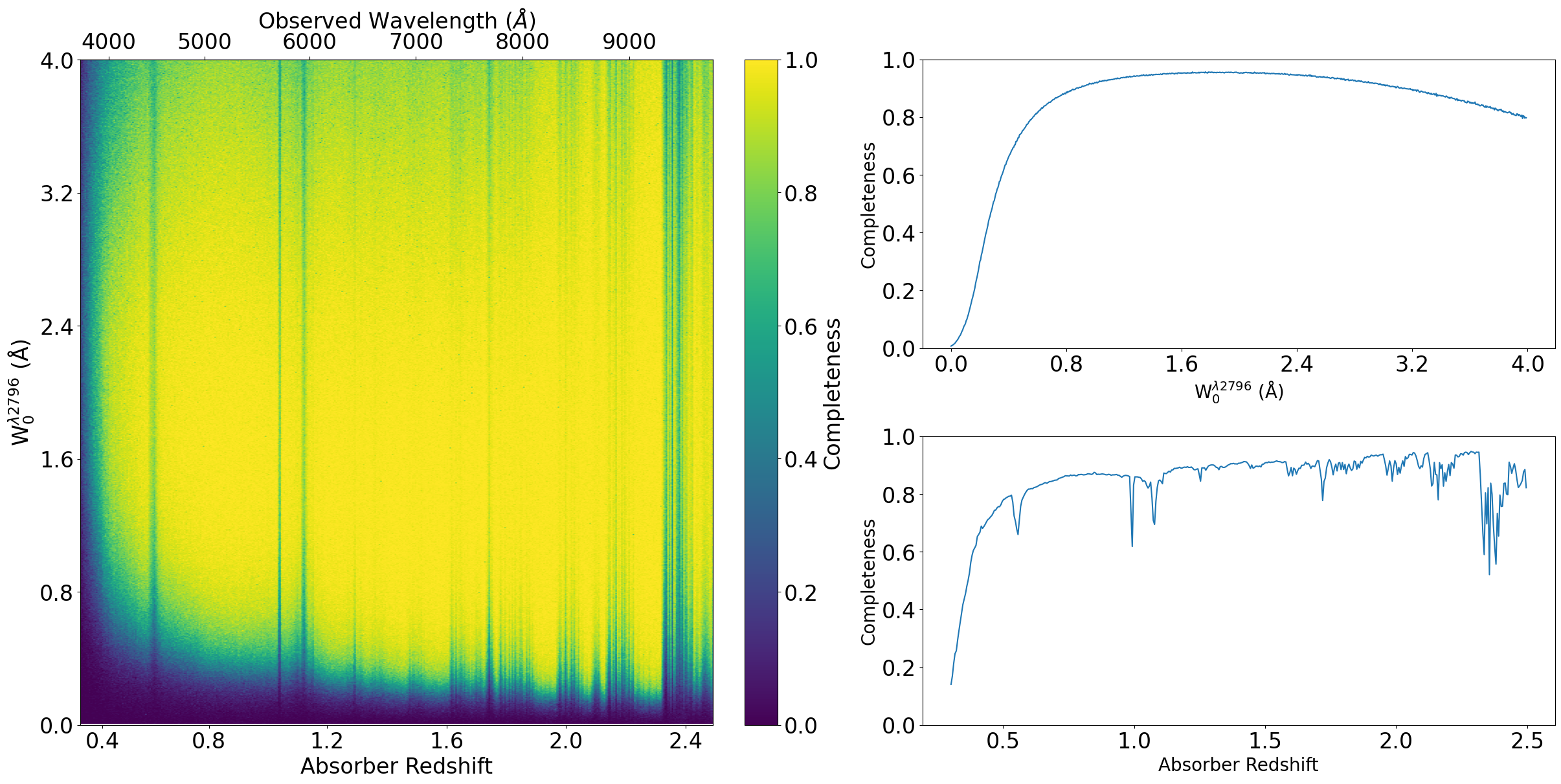}
    \caption{\textit{Left}: Heatmap of synthetic absorber recovery results. Decreases in completeness at low values of W$_0^{\lambda 2796}$ as well as at low redshift, and at certain redshifts corresponding to known skylines can clearly be seen.  \textit{Top Right}: Average completeness as a function of W$_0^{\lambda 2796}$ \textit{Bottom Right}: Average completeness as a function of absorber redshift.}
    \label{EWsim}
\end{figure*}

In addition to determining our completeness with respect to the quality of searched spectra, it is also informative to determine our ability to recover absorbers with different innate absorption strengths, described by the rest-frame equivalent width of the 2796\AA\ line (W$_0^{\lambda 2796}$). In order to do so, we follow the approach of \citet{Zhu2013} (see their \S 3.3 and Figure 7) and generate synthetic absorber systems at a wide range of values of W$_0^{\lambda 2796}$.

These mock absorbers are generated by first setting W$_0^{\lambda 2796}$ equal to a random value between 0\AA\ and 4\AA. This upper limit has been chosen as systems with W$_0^{\lambda 2796} >$ 4\AA\ account for less than 3\% of our sample. The rest-frame equivalent width of the second line of the doublet, W$_0^{\lambda 2803}$, is then determined by drawing a value from the posterior of doublet ratio values (W$_0^{\lambda 2796}$ / W$_0^{\lambda 2803}$) found in our sample. In the same fashion we determine a value for the rest-frame Gaussian standard-deviation of both lines. This sets the line amplitudes, such that the determined values of W$_0$ are representative.

We have chosen to inject these systems in real DESI quasar spectra, specifically those quasars from the parent catalog used in this study in which no real \Mgii absorbers were detected. This ensures these spectra have realistic error properties. For quasars with $z < 2.15$, synthetic absorbers are injected at every pixel that corresponds to an absorber redshift between $z=0.3$ and $z=2.5$. For quasars with $z > 2.15$, in order to avoid the \LyA forest region, injection starts at the first pixel red-ward of the \LyA emission line.

After constructing the absorber model, as detailed above, the model is resampled into the DESI wavelength coverage. The flux values of pixels that lie within the absorption lines are then replaced by the model values. The model is then scaled by the average flux values in the pixels it is replacing, and noise is added, with values being drawn from a normal distribution centered on the error spectrum values of the pixels being replaced. The resulting spectrum is then run through the doublet-finder portion of our pipeline to test whether the injected absorber is recovered.

The results of this test, for a sample of 6000 randomly selected quasars, are shown in Figure \ref{EWsim}. The heatmap at the left clearly shows that our completeness decreases rapidly for systems below W$_0^{\lambda 2796}$ = 0.8 \AA. Also visible are regions of decreased completeness at constant absorber redshift. These are associated with common skylines, such as the high pressure sodium bump at $\sim5900$\AA\ or OH lines at $\sim 9200$\AA\ \citep[e.g.][]{Zhu2013}.

The marginalized distributions shown at the right of Figure \ref{EWsim} shows that in addition to decreasing rapidly at low values of W$_0^{\lambda 2796}$, our completeness also slowly decreases at W$_0^{\lambda 2796} > 2.4$\AA. We suspect this may be a result of the lines of these very strong absorbers blending together, such that they cannot be separately resolved by our code. We will seek to address this issue in future catalog releases.

Considering the completeness with respect to redshift, we can see that it slowly increases with redshift, and is generally lower at the far red and blue ends of the wavelength coverage. Dips in completeness at $z\sim1.1$ and $z\sim1.7$ are likely related to the DESI spectrograph crossover regions \citep{DESIInstrument}, where noise values tend to be higher. Completeness increasing with redshift is a notable difference when compared to the results of \citet{Zhu2013} and demonstrates that absorber searches performed with DESI will be more complete to absorbers at higher redshift because of its wavelength coverage and the depth of its observations.

It will be necessary to re-evaluate our completeness when considering observations taken during the DESI Main Survey, as direct tracer quasars with redshifts $1.6 < z < 2.15$ are unlikely to have been observed for four times the typical exposure time, as they were for survey validation observations.

Having outlined how the completeness of our sample is related to the signal-to-noise properties of the spectra analyzed, as well as the redshift and absorbing strength of the intervening systems, we will now consider the properties of the systems that were detected.
\section{Results}

From an initial sample of \ParentQSO quasars, we find a total sample of \PrePICut probable \Mgii systems. Following the cuts described in \S2.3, we reduce this sample to a total of \FinalSample physically possible \Mgii systems in 16{,}707 unique quasar spectra. This suggests that at least one absorber is detected in 20.1\% of quasar spectra and the overall occurrence rate of absorbers considering multi-absorber systems is 28.8\%. These results are in reasonable agreement with similar studies using SDSS data, which found at least one absorber in 10-20\% of quasar spectra \citep[e.g.][]{York2006,Raghunathan2016,Anand2021}. In this section we will consider the statistics of this sample as well as describe the structure of the catalog we generate from them.

\begin{figure}[htb!]
    \epsscale{1.15}
    \plotone{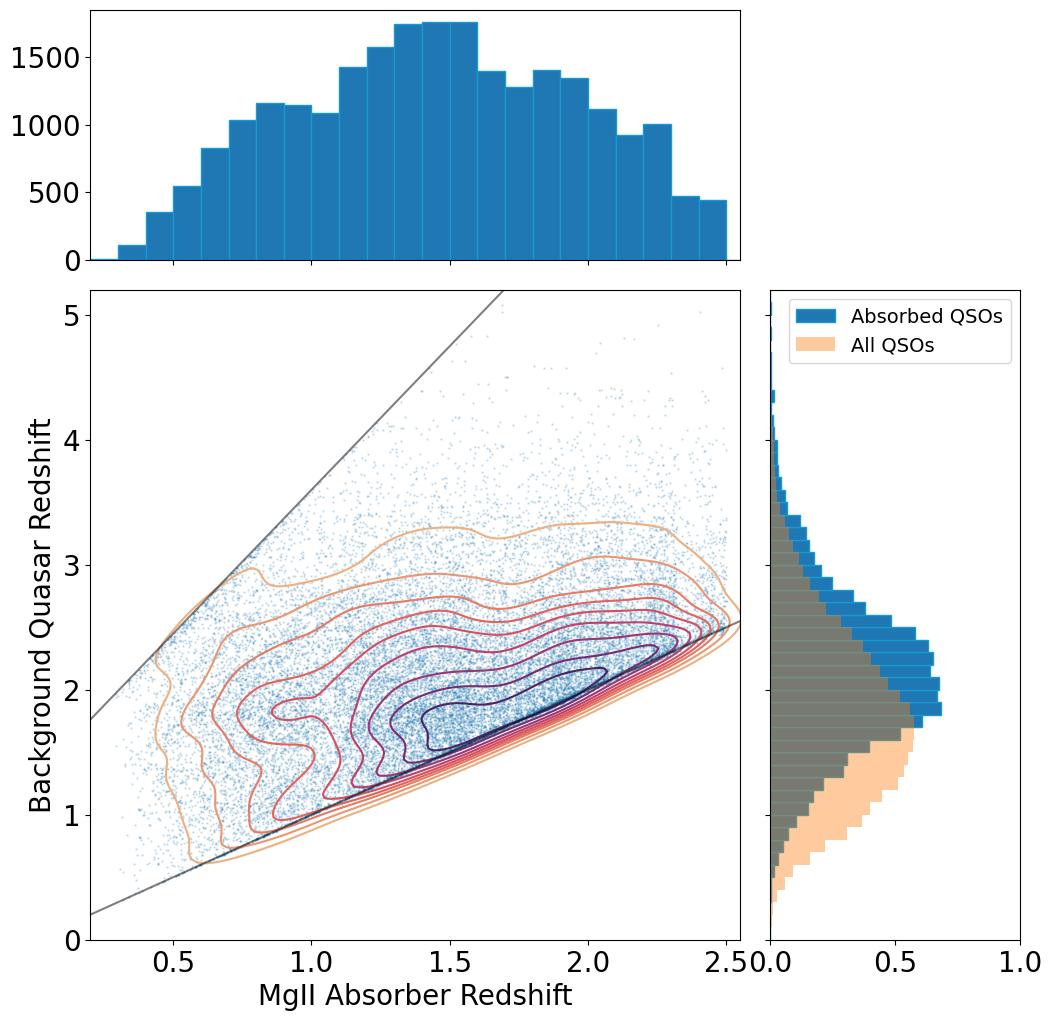}
    \caption{Redshift-space distributions of all detected absorbers. Contours are kernel density estimates of the distribution. Marginalized histograms of each redshift population are also presented. The quasar redshift histogram is plotted alongside the redshift histogram of all DESI QSOs for reference, and both are scaled by density.}
    \label{Redshifts}
\end{figure}

Figure \ref{Redshifts} displays the distribution of all detected absorption systems in both background quasar and absorber redshift-space. The overlaid contour lines are kernel density estimates and span 10\% to 90\% of the distribution in steps of 10\%. The overlaid black lines represent two natural ``boundaries'' for \Mgii systems. The lower boundary indicates the associated absorber case, in which the redshift of the quasar and absorption system are similar. As previously discussed this suggests that the absorption is occurring within the same galaxy, or galaxy cluster, that is host to the quasar. The upper boundary indicates the redshift of an absorber that would correspond to the wavelength of a quasar's \LyA emission line. As discussed in \S2 we exclude this region of redshift-space from our search due to contamination by the \LyA Forest. 

Figure~\ref{Redshifts} also presents marginalized histograms of the quasar and absorber redshifts. We can observe that the absorber redshift distribution is peaked between redshifts of 1.3 and 1.5 and declines in both directions from this peak. The precise physical interpretation of this histogram is complicated by redshift selection effects that are not yet well-characterized for the DESI survey, coupled with the true quasar and galaxy redshift distribution functions. 

The background quasar redshift distribution peaks in the redshift range $2.0 < z <  2.4$. This is in disagreement with the general DESI quasar distribution which peaks around $z=1.7$. Here a likely physical interpretation is that the likelihood of passing through an absorbing cloud is smaller for shorter lines of sight and therefore absorption is more likely to be found in higher redshift quasars. The decline at $z\,\,\geqsim 2.4$ is likely due to the overall reduction in the density of the quasar population at higher redshifts.

\begin{figure}[htb!]
    \epsscale{1.15}
    \plotone{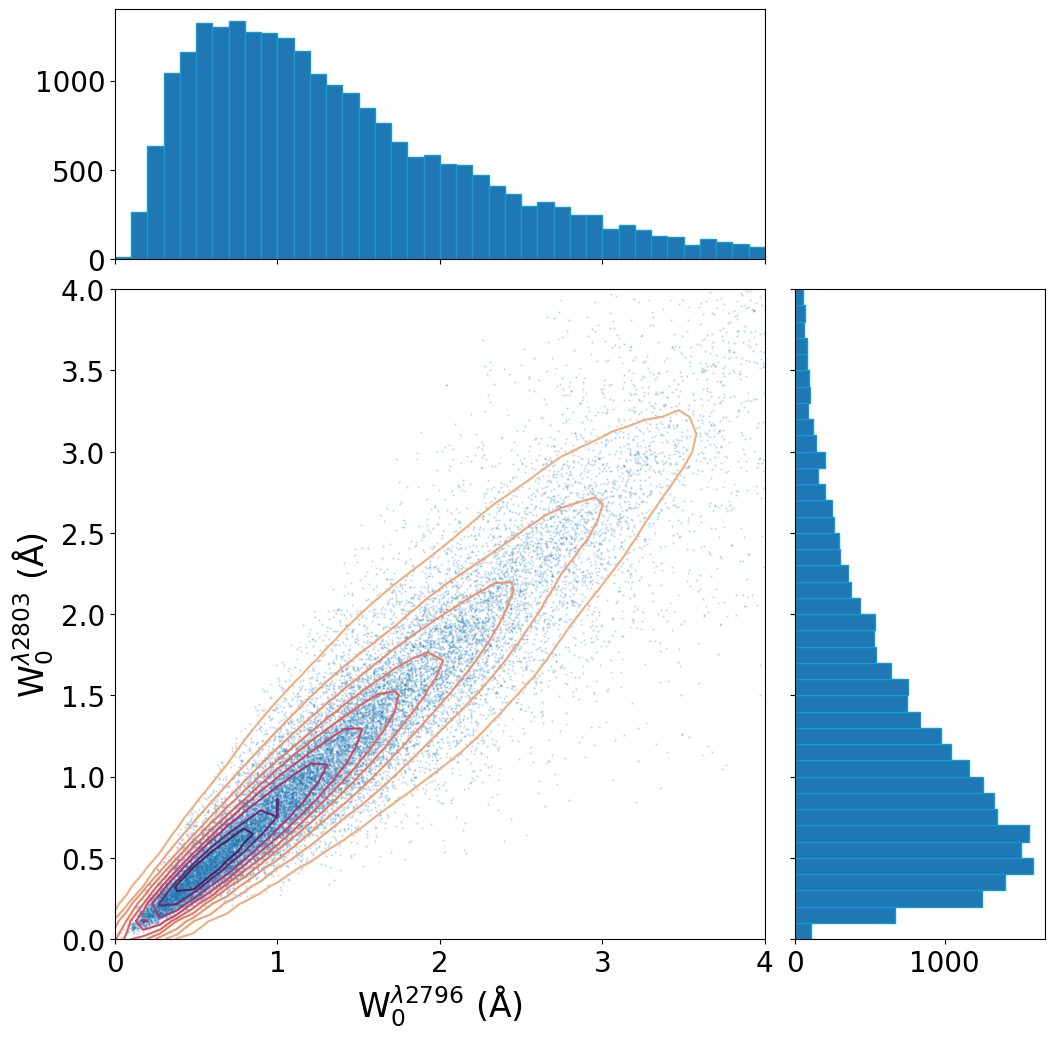}
    \caption{Distribution of rest-frame equivalent width values for the (2796\,\AA) and (2803\,\AA) \Mgii lines. Contours are kernel density estimates of the distribution. Marginalized histograms are also presented.}
    \label{EWs}
\end{figure}

Figure \ref{EWs} presents the distribution of measured rest-frame equivalent widths (EWs) for both lines of the \Mgii doublet. The overlaid contour lines follow the same scheme as in Figure \ref{Redshifts}. We observe that the region of highest density corresponds to absorbers with EWs between $\sim0.4$ to 1.0 \AA. Additionally, we note a slight skew to the contours, suggesting that the leading, 2796 \AA, line of the \Mgii doublet generally has a larger equivalent width value. This result is to be expected given the underlying atomic physics as discussed in \S \ref{sec:VI}.

In the marginalized distributions of W$_0$ values, populations of detected systems generally increase with decreasing W$_0$, until W$_0\sim0.8\AA$ at which point they flatten and then decrease rapidly at values $<0.4\AA$. Considering this alongside the completeness relationship we determined in \S \ref{sec:SynSim} we can infer that true absorber population sizes continue to increase with decreasing W$_0$, as has been seen in previous studies.

\subsection{Catalog Format and Access}
We describe the data columns that we catalog for each detected absorption system in the Appendix. We will now give a brief overview of these columns and their use. 

We first retain sufficient information to uniquely identify each analyzed healpix coadded spectrum, specifically the DESI {\tt TARGETID} (see \citealt{TargetingPaper}), R.A., decl., and phase of the DESI survey. We additionally store the {\tt Redrock} {\tt ZWARN} bitmask which details possible redshift warnings, as well as various TSNR2 values of each spectrum, and the best quasar redshift for each spectrum (derived from the parent quasar catalogs, as discussed in \S\ref{sec:pipeconstruct}). 

We also record the rest-frame EWs of both lines, as well as the central posterior distribution values for all five MCMC fit parameters, described in \S\ref{sec:pipeconstruct}. For the equivalent widths, as well as the fit parameters we also provide lower and upper error bars, as determined from the 16th and 84th percentiles of their posterior distributions. 

The created \Mgii absorber catalogs are available online\footnote{https://data.desi.lbl.gov/public/edr/vac/edr/mgii-absorber/v1.0/}. We have also retained the full 14,000-step MCMC chains for each detected absorber --- these will be made available upon request.

\section{Discussion}
In this section, we will consider the applications of our secondary catalog composed of physically impossible absorption systems. We will also examine the possibility of using \Mgii systems to detect other metal lines.

\subsection{Physically Impossible Absorbers}
\label{sec:PIAs}
\begin{figure}[htb!]
    \epsscale{1.17}
    \plotone{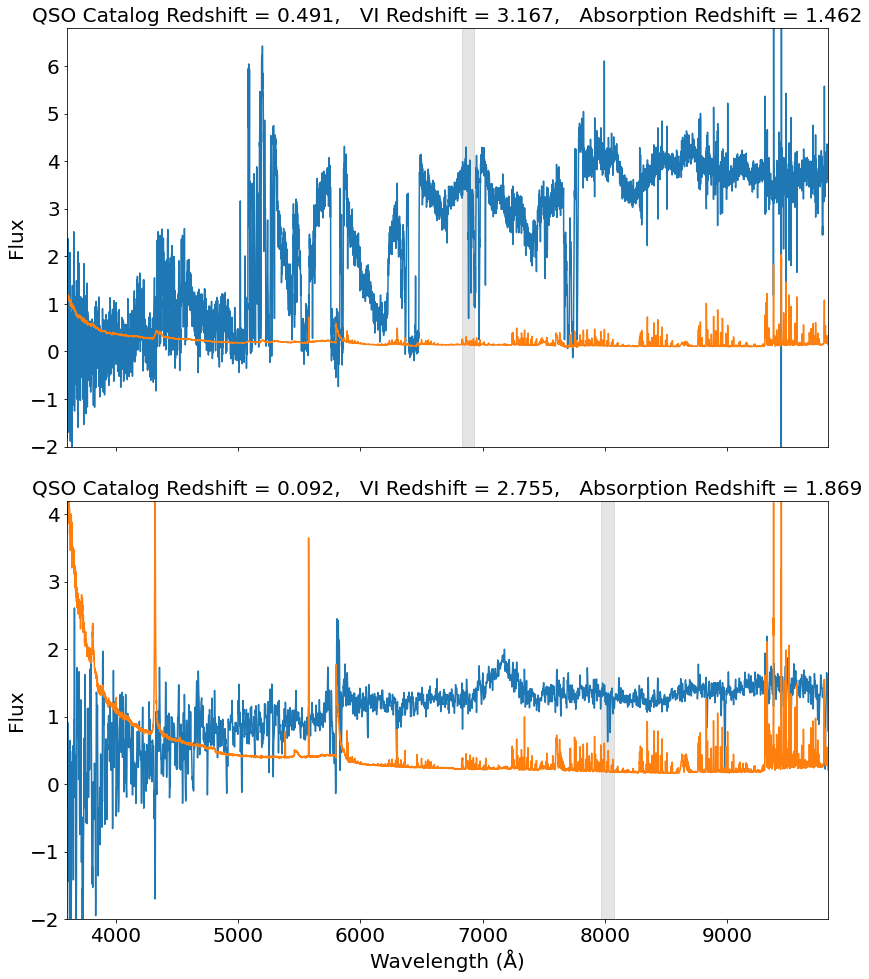}
    \caption{Two example spectra showing genuine \Mgii absorption systems that appear physically impossible due to poor pipeline redshifts. The detected \Mgii absorption is indicated by the grey outlined area. The associated error spectrum is shown in orange.}
    \label{PIA_ex}
\end{figure}

As discussed in \S2.2 we have detected a small number of systems that have an offset velocity that suggests the absorber is {\em behind} the quasar, which is physically impossible. In total there are \PISample such systems, which we will refer to as ``PI'' absorbers. These PI systems comprise roughly 1.3\% of our initial pre-quality-cuts sample of \PrePICut absorbers. To improve the utility of this PI subset we first apply the same quality cuts as for the main sample, which reduces the PI absorber sample to 108 systems in 84 unique spectra. Inspecting these spectra, we find 34 entries where the \Mgii absorption is clearly real, and the QSO redshift poorly determined. Nineteen of these systems are \LyA forest quasars and two illustrative spectra are shown in Figure \ref{PIA_ex}. We additionally find two instances of misidentified star spectra, and one instance of a QSO that has been redshifted at a value greater than that which we find in visual inspection; note that in these three cases the \Mgii absorption is not real.

The relatively low number of true PI absorbers that are found demonstrates the extremely high accuracy of the DESI redshift schema. Extrapolating from these results to the full five-year DESI sample we would anticipate finding only ~1200 true PI absorbers. Given these numbers it may be worthwhile to occasionally visually inspect these systems and reclassify any \LyA forest quasars with true PI absorbers such that they can be re-observed to improve the signal of the observation. We leave this consideration to future work.

\subsection{Detection of Additional Metal Lines}
\label{sec:metals}
Once an \Mgii absorption system has been identified at a particular redshift, we can search for other common metal lines, such as \Feiinospace, \CIVnospace, and \SiIVnospace, knowing precisely where in the spectrum these lines should appear. This enables the detection of these lines at relatively lower signal-to-noise.

A pilot analysis which involved the visual inspection of 1000 randomly selected \Mgii absorbers to search for \Feiinospace, \CIVnospace, and \SiIV at the same redshift yielded the results in Table~\ref{tab:abslines}. The ``Id.\ Rate'' column in Table~\ref{tab:abslines} shows the raw percentage of inspected \Mgii absorption systems in which the additional line could be identified, the ``Vis.\ Rate'' column shows the percentage of inspected absorption systems in which the non-\Mgii absorption species would be found at a wavelength $>$ 4000Å such that it would be readily visible to the DESI instrument.\footnote{I.e.\ as \SiIV has a wavelength of $\sim$1394\,\AA\ the Vis.\ Rate would be equal to the fraction of absorption systems at $z\,\,\geqsim 1.87$.} Finally, the ``Scaled Id.\ Rate'' column scales the Id.\ Rate by $[100 / \textrm{Vis.\ Rate}]$ to give the percentage of the time the line was identified when expected to be visible. We choose to use 4000\,\AA\ as this tends to be the region where the noise in DESI spectra reaches a consistent level (being noisier at lower wavelengths).

\begin{table}[t]
\centering
\caption{Additional Absorption Line VI Results}
\hspace{-45pt}
\begin{tabular}{c|c|c|c}
\hline
Line & Id. Rate & Vis.\ Rate & Scaled Id.\ Rate  \\
    \hline
    \hline
    \Feii  & 83.2\% & 91.3\% & 91.1\% \\
    \CIV & 27.9\% & 38.3\% & 72.8\% \\
    \SiIV & 16.2\% & 24.4\% & 66.4\% \\
\label{tab:abslines}
\end{tabular}
\end{table}

These results imply that (when visible) \Feii is identifiable in 91.1\% of systems and \CIV and \SiIV are similarly identifiable in 72.8\% and 66.4\% of systems, respectively. These results are promising and suggest that an algorithmic approach could reliably characterize additional absorption features when seeded with a redshift derived from a certain absorption doublet such as \Mgiinospace. 

\section{Conclusion}
In this paper we have presented the methods by which we detect and verify a sample of \Mgii absorption systems in the data collected during the Survey Validation phase of the DESI survey. We model these absorption systems by first identifying possible systems in smoothed residuals and then characterizing them using Gaussian lines profiles in an MCMC process. In total we have characterized \FinalSample absorption systems in 16{,}707 unique quasar spectra. The parent quasar catalogs utilized in this study contain \ParentQSO entries, implying that 20.1\% of DESI quasars will contain an identifiable \Mgii absorber. The total number of expected identifiable absorbers will then be equal to 28.8\% the total number of observed quasars (accounting for spectra with multiple absorbers). Assuming that DESI ultimately obtains spectra for 3 million quasars, our pilot study implies that DESI will eventually compile a sample of over 800{,}000 \Mgii absorption systems across $\sim$560{,}000 quasar spectra --- by far the largest such sample ever constructed.

The statistics of this catalog, a 99.1\% purity and 82.6\% completeness have been verified through the visual inspection of a subset of absorbers as well as the reanalysis of lower signal-to-noise spectra of objects for which absorption systems have been detected.

The estimated purity of this catalog (99.1\%), as well as the various factors affecting its completeness have been characterized through the visual inspection of a subset of absorbers, as well as the reanalysis of lower signal-to-noise spectra, and the construction of mock \Mgii absorbers. In future catalog releases we will aim to increase the completeness of this sample, either by reducing the doublet signal-to-noise threshold or by introducing an additional detection step that can recover \Mgii absorbers that currently escape detection due to high noise or unusual features, such as the high W$_0^{\lambda 2796}$ systems potentially missed due to line blending. This goal will of course require the careful balancing of completeness and purity --- for this first catalog release we have chosen to favor a catalog with high purity.

Additionally, we have made the choice at this time to group absorbers that appear to have physically impossible redshifts --- i.e.\ those that would suggest the absorption system to be farther from the observer than the quasar --- into a separate catalog. Such systems account for roughly 1.3\% of detected absorbers. From visual inspection of such systems we find that after applying the purity cuts described in \S2.3 around 40\% of these systems are true \Mgii absorbers with incorrect background quasar redshifts. We anticipate exploring the possibility of using these systems to improve DESI quasar redshifts.

We detect \Mgii absorbers in the redshift range $0.3\,\,\leqsim z\,\, \leqsim 2.5$ with a peak in the distribution of absorbers between $z\sim1.3$ and $z\sim1.5$. The exact interpretation of the redshift distribution of possible absorbers is difficult to disentangle from various selection effects. The background quasars which enable the observations of these systems are found at $0.4\,\,\leqsim z\,\, \leqsim 5.8$ and, as can be seen in Figure \ref{Redshifts}, are generally at higher redshifts than the full DESI quasar population. 

The physical properties of the absorption systems catalogued here, such as relative metallicities, ionization temperatures, and physical densities, can be determined by further analysis. In order to do so we plan to automate the detection and characterization of additional metal lines. As noted in \S\ref{sec:metals}, we identify at least one additional line in $>91.1$\% of \Mgii absorbers. The equivalent widths of the \Mgii absorption systems discussed in this paper are generally similar between the two lines of the \Mgii doublet and can be found at levels well below 1Å, suggesting that even weak absorbers can be readily detected.

The sample of absorbers collected here already is sufficiently large to facilitate a variety of studies including; the nature of the CGM environments from which these absorption systems arise, the clustering of underlying dark matter traced by the three dimensional locations of the absorbers, or the use of these \Mgii systems to find additional species in absorption.

\begin{acknowledgments}
We thank Guangtun Zhu for sharing the NMF eigen-spectra of SDSS quasars. We thank the referee for their valuable feedback that has helped us improve the quality of the paper.

LN and ADM were supported by the U.S.\ Department of Energy, Office of Science, Office of High Energy Physics, under Award Number DE-SC0019022. LN was also partially supported by Wyoming NASA Space Grant Consortium award \#80NSSC20M0113. AP was supported by the University of Wyoming Science Initiative Wyoming Research Scholars Program. TWL was supported by the Ministry of Science and Technology (MOST 111-2112-M-002-015-MY3), the Ministry of Education, Taiwan (MOE Yushan Young Scholar grant NTU-110VV007), National Taiwan University research grants (NTU-CC-111L894806, NTU-111L7318). 

This research is supported by the Director, Office of Science, Office of High Energy Physics of the U.S. Department of Energy under Contract No. DE–AC02–05CH11231, and by the National Energy Research Scientific Computing Center, a DOE Office of Science User Facility under the same contract; additional support for DESI is provided by the U.S. National Science Foundation, Division of Astronomical Sciences under Contract No. AST-0950945 to the NSF’s National Optical-Infrared Astronomy Research Laboratory; the Science and Technologies Facilities Council of the United Kingdom; the Gordon and Betty Moore Foundation; the Heising-Simons Foundation; the French Alternative Energies and Atomic Energy Commission (CEA); the National Council of Science and Technology of Mexico (CONACYT); the Ministry of Science and Innovation of Spain (MICINN), and by the DESI Member Institutions: \url{https://www.desi.lbl.gov/collaborating-institutions}.

The authors are honored to be permitted to conduct scientific research on Iolkam Du’ag (Kitt Peak), a mountain with particular significance to the Tohono O’odham Nation.

The data used to construct Figures 1, 2, 6, 7, and 8 is available in Zenodo at \dataset[DOI: 10.5281/zenodo.7826591]{https://doi.org/10.5281/zenodo.7826591}

\end{acknowledgments}

\bibliographystyle{yahapj}
\bibliography{MgII.bib}

%\clearpage
\appendix
We provide two catalogs containing the results of our \Mgii absorber search. Both catalogs share the common format denoted in the table below.

\textbf{MgII-Absorbers-EDR.fits}: Primary catalog containing \FinalSample identified absorbers that appear physically possible, i.e. have velocity offsets less than 5000 ${\rm km\,s}^{-1}$ relative to the QSO redshift.

\textbf{MgII-Absorbers-EDR-PI.fits}: Secondary catalog containing 108 identified absorbers that appear physically impossible, i.e. have velocity offsets greater than 5000 ${\rm km\,s}^{-1}$ relative to the QSO redshift.

\section{Catalog Format}\label{asec:append}
\startlongtable
\begin{deluxetable*}{clcl}
\tabletypesize{\footnotesize}
\tablecaption{Format of the MgII absorber catalog$^a$ \label{tab:fields}}
\tablecolumns{4}
\tablewidth{0pt}
\tablehead{
\colhead{Column} & \colhead{Name} & \colhead{Format} & \colhead{Description}
}
\startdata
1 & TARGETID & INT & Unique DESI Target Designation\\
2 & RA & DOUBLE & Right Ascension in decimal degrees (J2000)\\
3 & DEC & DOUBLE & Declination in decimal degrees (J2000)\\
4 & SURVEY & STR & Survey validation stage (sv1, sv2, or sv3) of analyzed spectrum \\
5 & ZWARN & INT & Redrock redshift warning bitmask$^b$ \\
6 & TSNR2\_QSO & DOUBLE & Quasar template signal-to-noise value squared \\
7 & TSNR2\_LYA & DOUBLE & Lyman-$\alpha$ quasar template signal-to-noise value squared \\
8 & TSNR2\_LRG & DOUBLE & Luminous Red Galaxy template signal-to-noise value squared \\
9 & Z\_QSO & DOUBLE & Redshift value from input quasar catalog\\
\hline
10 & EW\_2796 & DOUBLE & Central posterior value for W$_0^{\lambda 2796} (\textrm{\AA})$ \\
11 & EW\_2803 & DOUBLE & Central posterior value for W$_0^{\lambda 2803} (\textrm{\AA})$ \\
12 & EW\_2796\_ERR\_LOW & DOUBLE & Lower error bar for W$_0^{\lambda 2796} (\textrm{\AA})$ \\
13 & EW\_2803\_ERR\_LOW & DOUBLE & Lower error bar for W$_0^{\lambda 2803} (\textrm{\AA})$ \\
14 & EW\_2796\_ERR\_HIGH & DOUBLE & Upper error bar for W$_0^{\lambda 2796} (\textrm{\AA})$ \\
15 & EW\_2803\_ERR\_HIGH & DOUBLE & Upper error bar for W$_0^{\lambda 2803} (\textrm{\AA})$ \\
\hline
16 & Z\_MGII & DOUBLE & Central posterior value for the redshift of the Mg II absorption feature \\
17 & AMP\_2796 & DOUBLE & Central posterior value for the amplitude of the 2796$\textrm{\AA}$ line \\
18 & AMP\_2803 & DOUBLE & Central posterior value for the amplitude of the 2803$\textrm{\AA}$ line \\
19 & STDDEV\_2796 & DOUBLE & Central posterior value for the standard deviation of the 2796$\textrm{\AA}$ line \\
20 & STDDEV\_2803 & DOUBLE & Central posterior value for the standard deviation of the 2803$\textrm{\AA}$ line \\
\hline
21 & Z\_MGII\_ERR\_LOW & DOUBLE & Lower error bar for the redshift of the Mg II absorption feature \\
22 & AMP\_2796\_ERR\_LOW & DOUBLE & Lower error bar for the amplitude of the 2796$\textrm{\AA}$ line \\
23 & AMP\_2803\_ERR\_LOW & DOUBLE & Lower error bar for the amplitude of the 2803$\textrm{\AA}$ line \\
24 & STDDEV\_2796\_ERR\_LOW & DOUBLE & Lower error bar for the standard deviation of the 2796$\textrm{\AA}$ line \\
25 & STDDEV\_2803\_ERR\_LOW & DOUBLE & Lower error bar for the standard deviation of the 2803$\textrm{\AA}$ line \\
\hline
26 & Z\_MGII\_ERR\_HIGH & DOUBLE & Upper error bar for the redshift of the Mg II absorption feature \\
27 & AMP\_2796\_ERR\_HIGH & DOUBLE & Upper error bar for the amplitude of the 2796$\textrm{\AA}$ line \\
28 & AMP\_2803\_ERR\_HIGH & DOUBLE & Upper error bar for the amplitude of the 2803$\textrm{\AA}$ line \\
29 & STDDEV\_2796\_ERR\_HIGH & DOUBLE & Upper error bar for the standard deviation of the 2796$\textrm{\AA}$ line \\
30 & STDDEV\_2803\_ERR\_HIGH & DOUBLE & Upper error bar for the standard deviation of the 2803$\textrm{\AA}$ line \\
\hline
31 & CONTINUUM\_METHOD & STR & Method by which QSO continuum was determined while fitting ('NMF' or 'Medianfilter') \\
32 & LINE\_SNR\_MIN & DOUBLE &  Minimum SNR value of the two Mg II lines, used in initial detection \\
33 & LINE\_SNR\_MAX & DOUBLE &  Maximum SNR value of the two Mg II lines, used in initial detection \\
\hline
\enddata
\end{deluxetable*}

\onecolumngrid

\footnotesize{$^b$ Documented at https://github.com/desihub/redrock/blob/0.17.0/py/redrock/zwarning.py}

\(\) %Text after forces full table to be displayed.

\end{document}